\documentclass[notitlepage,a4paper,aps,prd,onecolumn,superscriptaddress,nofootinbib,groupedaddress]{revtex4}

\usepackage{amsmath,comment}
\usepackage{amsfonts,color}
\usepackage{amsmath}
\usepackage{amsthm}
\usepackage{pdfpages}
\usepackage{amsmath}
\usepackage{empheq}
\usepackage{amsfonts,color}
\usepackage{amssymb}
\usepackage{physics}
\usepackage{booktabs,multirow}

\usepackage[utf8]{inputenc}
\allowdisplaybreaks
\usepackage{color}
\usepackage{hyperref}
\hypersetup{
    colorlinks=true,
    linkcolor=blue,
    filecolor=magenta,      
   citecolor=blue
}

\usepackage{enumitem}

\usepackage{tikz}
\usetikzlibrary{shapes.geometric}
\usetikzlibrary{arrows.meta,arrows}

\begin{document}


\title{Coupling Electromagnetism to Torsion: Black Holes and Spin-Charge Interactions}

\author{Sebastian Bahamonde}
\email{sbahamondebeltran@gmail.com, sebastian.bahamonde@ipmu.jp}
\affiliation{Kavli Institute for the Physics and Mathematics of the Universe (WPI), The University of Tokyo Institutes
for Advanced Study (UTIAS), The University of Tokyo, Kashiwa, Chiba 277-8583, Japan.}
\affiliation{Cosmology, Gravity, and Astroparticle Physics Group, Center for Theoretical Physics of the Universe,
Institute for Basic Science (IBS), Daejeon, 34126, Korea.}

\author{Jorge Maggiolo}
\email{jorge.maggiolo.t@gmail.com}
\affiliation{Pontificia Universidad Católica de Valparaíso, Instituto de Física, Avenida Brasil 2950, Valparaíso, Chile.}
\affiliation{Departamento de Física, Universidad Técnica Federico Santa María, Casilla 110-V, Valparaíso, Chile.}

\author{Christian Pfeifer}
\email{christian.pfeifer@zarm.uni-bremen.de}
\affiliation{ZARM, University of Bremen, 28359 Bremen, Germany.}

\begin{abstract}
The coupling between matter fields and gravity (encoded in the geometry of spacetime) can be realized in various ways. Most commonly, a minimal coupling principle is employed, meaning that all matter fields, except spinors, couple only to the spacetime metric, while spinors additionally couple to the spacetime connection. Non-minimal couplings between matter fields and spacetime curvature can arise, for example, from quantum field theory on curved spacetime through renormalization corrections, in gauge theories of gravity, and in effective field theories. In this article, we consider a non-minimal coupling $F^{\mu\nu}\tilde{R}_{\mu\nu}$ between the field strength tensor of the electromagnetic field $F_{\mu\nu}$ and the antisymmetric part of the Ricci tensor $\tilde{R}_{[\mu\nu]}$ in Riemann–Cartan geometry, which is based on a general metric-compatible connection with torsion. We find an exact 4-dimensional vacuum solution that generalizes the Reissner–Nordström black hole from Einstein-Maxwell and reveals new interactions between the intrinsic torsion-spin charge and the electric charge. Qualitatively, this solution exhibits two distinct features: the effective charge is not constrained to be positive, and the sign of the electric charge influences its gravitational effects. We also derive slowly rotating solutions in 3 dimensions, representing a generalized slowly rotating BTZ black hole solution with couplings among the magnetic and electric charges, the angular momentum, and the intrinsic torsion-spin charge.
\end{abstract}

\maketitle

\section{Introduction}\label{sec:intro}
Since the formulation of General Relativity (GR), we describe the gravitational interaction through the geometry of spacetime. The simplest version, on which GR is based, is pseudo-Riemannian geometry, where only a spacetime metric (and its metric-compatible, torsion free Levi-Civita connection) encode the gravitational degrees of freedom.

Clearly, pseudo-Riemannian geometry is not the only possibility to describe the geometry of spacetime. On the one hand one can consider the affine connection structure independently from the metric, including properties like torsion \cite{Hehl:1976kj,Bahamonde:2021gfp} and non-metricity \cite{Hehl:1994ue,JimenezCano:2021rlu}. On the other hand, one could consider non-linear connections as in Finsler \cite{Heefer:2024kfi,Pfeifer:2019wus} or Hamilton geometry \cite{Barcaroli:2015xda}. Since GR and pseudo-Riemmanian geometry together with the standard model or particle physics might not be our final understanding of the gravity, as for example, can be seen from the observational tensions in cosmology \cite{CosmoVerse:2025txj} and the missing theory of quantum gravity \cite{Addazi:2021xuf}, spacetime geometries beyond pseudo-Riemannian geometries are investigated as viable extensions of GR. An important aspect in this research is to understand the coupling between matter fields and gravity, i.e.\ between matter fields and the geometry of spacetime.

In GR, together with the standard model of particle physics, the metric minimal coupling principle is employed, meaning that all matter fields only couple to the spacetime metric, and spinors couple in addition to the Levi-Civita connection. In the framework of quantum field theory on curved spacetime, couplings between curvature and matter fields appear from renormalization corrections to the original action, for example when one considers quantum electrodynamics on curved spacetime \cite{DrummondHathrell}. It is expected that when one extends the geometry of spacetime beyond pseudo-Riemannian geometry, non-minimal couplings between matter fields and additional geometric fields emerge.

A particularly interesting class of extended spacetime geometries that allow for a broader interaction between spacetime geometry and matter fields, are Riemann-Cartan spacetime geometries. They are based on a spacetime metric and an independent, metric compatible connection with torsion and are the mathematical basis for the formulation of Poincar\'e gauge gravity theories~\cite{Hehl:1976kj,Hehl:1994ue,Blagojevic:2013xpa,Obukhov:2022khx}. Compared to pseudo-Riemannian spacetime geometry and GR, additional geometric degrees of freedom, describing gravity, are encoded in the torsion tensor. Due to its symmetries, it naturally couples to the electromagnetic field strength tensor, which is why a non-minimal coupling between torsion and electromagnetism has gained particular attention. For example, it has been studied in depth how torsion influences the propagation and structure of electric and magnetic fields~\cite{Hehl:1999bt,Itin:2003hr,deAndrade:1997cj,Balakin:2007am,Solanki:2004az, Hammond:1991ty,Delhom:2020hkb,Dias2005,Hehl:2016wwp}. These studies have demonstrated that torsion can induce anisotropic modifications in the propagation of electromagnetic waves, alter the effective field strength perceived by charged particles, and break the duality symmetry between electric and magnetic components, leading to direction-dependent corrections in electromagnetic interactions. Several works have proposed non-minimal couplings between torsion and the electromagnetic field within effective field theories~\cite{shapiro2002,HehlObukhov2003}, while others have interpreted certain components of the torsion tensor as potential geometric origins of electromagnetic phenomena~\cite{Hammond2002}. Furthermore, black hole solutions incorporating torsion have been developed in both metric-affine and gauge-theoretic formulations of gravity~\cite{Lenzen:1985ess,Tresguerres:1995js,Tresguerres:1995un,Vlachynsky:1996zh,Obukhov:1996pf,Ho:1997xbz,Obukhov:1997uc,Blagojevic2002,Baekler:2006de,Obukhov:2012je,Giribet:2014hpa,Cembranos:2016xqx,Cembranos:2016gdt,Cembranos:2017pcs,Obukhov:2019fti,Bahamonde:2020fnq,Bahamonde:2020vpb,Nikiforova:2020sac,Pfeifer:2021njm,DAmbrosio:2021zpm,Bahamonde:2020snl,Bahamonde:2021qjk,Bahamonde:2022jue,Bahamonde:2022esv,Bahamonde:2021srr,Bahamonde:2022lvh,Bahamonde:2022chq,Bahamonde:2022kwg,Bahamonde:2022meb,Gonzalez:2024ifp,Bahamonde:2024sqo}, highlighting the rich phenomenology that torsion introduces to classical gravitational configurations. Also, in lower-dimensional settings, three-dimensional gravity models with electric fields have been constructed and analysed, for example in~\cite{Blagojevic:2013bu, Mielke:1991nn,Mielke:2003xx,Garcia:2003nm,Blagojevic:2003wn}.

Interestingly, a simple and direct way to couple torsion and the electromagnetic field by terms that are linear in the electromagnetic field strength $F_{\mu\nu}$ has not been investigated in all detail yet. In particular, the simple term that is a product $\tilde{R}_{[\mu\nu]} F^{\mu\nu}$, between the electromagnetic field strength tensor and the antisymmetric part of the Ricci tensor $\tilde{R}_{[\mu\nu]}$ of the Riemann–Cartan connection, has not been thoroughly considered.

In this work, we write down the most general gravitational theory that consists of couplings that are at most linear linear in the electromagnetic field strength $F_{\mu\nu}$, linear in the Riemann-Cartan curvature $\tilde{R}^\lambda{}_{\rho\mu\nu}$, and quadratic in the torsion $T^\lambda{}_{\mu\nu}$. We mainly focus on the new coupling of the form $F^{\mu\nu}\tilde{R}_{\mu\nu}$, which arises naturally from the geometric structure, preserves the gauge symmetry of the electromagnetic sector, and maintains general covariance.

Having identified the action of interest, we derive the corresponding field equations in the tetrad/spin connection formulation of Riemann-Cartan theories of gravity, and explore two physically distinct settings. The first scenario involves a static, spherically symmetric configuration in four dimensions, where we find exact black hole solutions in which torsion plays a central role. Not only does it alter the geometry of spacetime, but it also modifies the effective electric charge through the emergence of an intrinsic spin charge that couples directly to the electric field—giving rise to corrections absent in the standard Reissner–Nordström framework.  For the second scenario, we turn to three-dimensional gravity with a negative cosmological constant, a setting that has been extensively explored in the context of generalizations of the BTZ black hole~\cite{Banados:1992wn,Banados:1992gq}, with torsion acting as a cosmological constant~\cite{Mielke:2003xx,Garcia:2003nm,Blagojevic:2003wn,Canfora:2007fw,Blagojevic:2008ip,Blagojevic:2008xz,Cvetkovic:2008zz,Vasquez:2009mk,Blagojevic:2009xw,Cvetkovic:2018ati,Canfora:2010rh,Blagojevic:2013aaa,Becar:2013qba,Gonzalez:2014pwa,Ma:2014tka,Aviles:2023igk}. For the theory under discussion we construct exact solutions in the regime of slow rotation and the torsion tensor acts as an electric potential. In the static limit, the coupling between spin and electric charge reflects the behaviour observed in four dimensions. Remarkably, once rotation is introduced perturbatively, a new term emerges that links the intrinsic spin charge to the angular momentum, an effect absent in the original charged BTZ geometry~\cite{Martinez:1999qi}. Altogether, our findings reveal new mechanisms through which torsion can influence gravitational and gauge dynamics, leading to nontrivial spacetime structures and couplings with physical observables such as electric charge and angular momentum.

We organise the presentation of our results as follows. In Sec.\ref{sec:conventions}, we present the basic concepts of Riemann-Cartan geometries. Then, in Sec.\ref{secelectro}, we construct and introduce our theory with new couplings between the electromagnetic field and the generalised Ricci tensor. In Secs.~\ref{sec:4D} and~\ref{sec:3D}, we present new black hole solutions in four and three dimensions, respectively. Finally, in Sec.~\ref{sec:conc}, we summarise the main conclusions and outline potential directions for future research within this framework. Throughout this article we use the following conventions: The signature of the metric is mostly minus $(+,-,...,-)$. A tilde $\tilde{}$ over an object denotes quantities computed with the general metric compatible connection with torsion.

\section{Riemann-Cartan geometries}\label{sec:conventions}
We start by briefly introducing the relevant tensorial quantities needed to write down the non-minimally coupled Riemann-Cartan-Maxwell theories in the next Sec.~\ref{secelectro}.

A Riemann-Cartan spacetime $(M,g,\tilde \Gamma)$ is an n-dimensional manifold $M$ equipped with a Lorentzian metric and a metric compatible connection $\tilde \Gamma$ with torsion \cite{Hehl:1994ue,Obukhov:2022khx}.

In local coordinates, the components of the torsion are given by,
\begin{align}
    T^{\lambda}\,_{\mu \nu}&=2\tilde{\Gamma}^{\lambda}\,_{[\mu \nu]}\,,
\end{align}
which allows us to decompose the $\tilde \Gamma$ covariant derivative $\tilde{\nabla}_{\mu}v^{\lambda}$ of an arbitrary vector $v^{\lambda}$ into a Levi-Civita covariant derivative $\nabla_{\mu}v^{\lambda}$ and the so-called contortion tensor $K^{\lambda}\,_{\rho\mu}$
\begin{align}
\tilde{\nabla}_{\mu}v^{\lambda}&=\nabla_{\mu}v^{\lambda}+K^{\lambda}\,_{\rho\mu}v^{\rho}
\end{align}
with
\begin{align}
    K^{\lambda}\,_{\rho\mu}&=\frac{1}{2}(T^{\lambda}\,_{\rho\mu}-T_{\rho}\,^{\lambda}\,_{\mu}-T_{\mu}\,^{\lambda}\,_{\rho})\,.\label{contortion}    
\end{align}
The general connection coefficients $\tilde\Gamma^\lambda{}_{\rho\mu}$ are decomposed into the Levi-Civita connection coefficients $\Gamma^\lambda{}_{\rho\mu}$, that are defined solely through the metric, and the contortion as
\begin{align}
   \tilde\Gamma^\lambda{}_{\rho\mu}=\Gamma^\lambda{}_{\rho\mu} +K^{\lambda}{}_{\rho\mu}\,.
\end{align}

The curvature $\tilde R$ of the connection $\tilde \Gamma$ can be obtained from the commutator of the covariant derivative as
\begin{equation}
[\tilde{\nabla}_{\mu},\tilde{\nabla}_{\nu}]\,v^{\lambda}=\tilde{R}^{\lambda}\,_{\rho \mu \nu}\,v^{\rho}+T^{\rho}\,_{\mu \nu}\,\tilde{\nabla}_{\rho}v^{\lambda}\,,
\end{equation}
where
\begin{equation}\label{totalcurvature}
\tilde{R}^{\lambda}\,_{\rho \mu \nu}=\partial_{\mu}\tilde{\Gamma}^{\lambda}\,_{\rho \nu}-\partial_{\nu}\tilde{\Gamma}^{\lambda}\,_{\rho \mu}+\tilde{\Gamma}^{\lambda}\,_{\sigma \mu}\tilde{\Gamma}^{\sigma}\,_{\rho \nu}-\tilde{\Gamma}^{\lambda}\,_{\sigma \nu}\tilde{\Gamma}^{\sigma}\,_{\rho \mu}\,.
\end{equation}
The corresponding Ricci tensor is given by
\begin{eqnarray}
    \tilde{R}_{\mu\nu}&=&\tilde{R}^{\lambda}\,_{\mu \lambda \nu}=  \tilde{R}_{(\mu\nu)}+  \tilde{R}_{[\mu\nu]}\,,\label{ricciT}
\end{eqnarray}
and in contrast to the pseudo-Riemannian case, is not necessarily symmetric. The symmetric and antisymmetric parts can be expressed through the symmetric pseudo-Riemannian Ricci tensor $R_{\mu\nu}$ and the torsion tensor as
\begin{eqnarray}
   \tilde{R}_{(\mu\nu)}&=&R_{\mu  \nu  } + \frac{1}{4} T_{\mu  }{}^{\alpha  \rho  } T_{\nu  \alpha  \rho  }+\frac{1}{2} T_{(\mu  }{}^{\alpha  \rho  }T_{|\alpha|  \nu)  \rho  }  + T^{\rho  }{}_{\alpha  \rho  } T_{(\mu  \nu)  }{}^{\alpha  } + \nabla_{\alpha  }T_{(\mu  \nu)  }{}^{\alpha  }  - \nabla_{(\mu  }T^{\alpha  }{}_{\nu)  \alpha  }\,,\\  
    \tilde{R}_{[\mu\nu]}&=&\frac{1}{2} T^{\alpha  }{}_{\mu  \nu  } T^{\rho  }{}_{\alpha  \rho  } + \frac{1}{2} \nabla_{\alpha  }T^{\alpha  }{}_{\mu  \nu  }+\frac{1}{2}  T_{[\mu  }{}^{\alpha  \rho  }T_{|\alpha | \nu]  \rho  } + \nabla_{[\mu  }T^{\alpha  }{}_{\nu]  \alpha  }\,.
\end{eqnarray}
It becomes clear that for vanishing torsion, the antisymmetric components vanish. The Riemann-Cartan Ricci scalar is given by the trace of the Ricci tensor:
\begin{eqnarray}\label{eq:ECRicciS}
    \tilde{R}=g^{\mu\nu}\tilde{R}_{\mu\nu} = R 
    + \frac{1}{4} T^{\mu\alpha  \rho  }T_{\mu  \alpha  \rho  }
    + \frac{1}{2} T^{\mu \alpha  \rho  }T_{\alpha  \mu  \rho  }  
    - T_\alpha T^{\alpha  } - 2 \nabla_{\alpha  }T^{\alpha  }\,,
\end{eqnarray}
where we introduced the vector torsion $T_\alpha = T^\mu{}_{\alpha\mu}$.

For constructing Riemann-Cartan theories of gravity and to introduce spinors, it is convenient to introduce tetrads  $e^a{}_\mu$ and a principal bundle connection $\tilde \omega_{\mu} \in \mathfrak{so}(1,3)$ that leads to a Poincar\'e gauge characterisation of the geometry of  spacetime \cite{Hehl:1976kj,ponomarev2017gauge}. The metric and the affine connection can be related to the tetrad and spin connection as
\begin{eqnarray}
g_{\mu \nu}&=&e^{a}\,_{\mu}\,e^{b}\,_{\nu}\,\eta_{a b}\,,\label{vierbein_def}\\
\tilde{\omega}^{a}\,_{b\mu}&=&e^{a}\,_{\lambda}\,e_{b}\,^{\rho}\,\tilde{\Gamma}^{\lambda}\,_{\rho \mu}+e^{a}\,_{\lambda}\,\partial_{\mu}\,e_{b}\,^{\lambda}\,,\label{anholonomic_connection}
\end{eqnarray}
where $\eta_{ab}$ is the Minwkoski metric. Then, one can define the torsion and curvature gauge field strengths (also called gauge curvatures) as
\begin{eqnarray}
    F^{a}\,_{\mu\nu}&=& \partial_{\mu}e^{a}\,_{\nu}-\partial_{\nu}e^{a}\,_{\mu}+\tilde\omega^{a}\,_{b\mu}\,e^{b}\,_{\nu}-\tilde\omega^{a}\,_{b\nu}\,e^{b}\,_{\mu} 
    = e^{a}\,_{\lambda}T^{\lambda}\,_{\nu\mu}\,,\label{F1}\\
    F^{a}\,_{b\mu\nu}&=&\partial_{\mu}\tilde\omega^{a}\,_{b\nu} -\partial_{\nu}\tilde\omega^{a}\,_{b\mu}+\tilde\omega^{a}\,_{c\mu}\,\tilde\omega^{c}\,_{b}\,_{\nu}-\tilde\omega^{a}\,_{c\nu}\,\tilde\omega^{c}\,_{b\mu}
    =\eta_{b c}\,e^{a}\,_{\lambda}e^{c\rho}\tilde{R}^{\lambda}\,_{\rho\mu\nu} \label{F2}\,.
\end{eqnarray}
This reformulation allows us to consider the tetrad and spin connection as fundamental variables of the theory, which is what is often done in the literature, and what also we will do in the following. 

In the following we will work mostly with spacetime indices $\{\mu,\nu, ...\}$. An object with tetrad indices $Z^a{}_b$ is converted to an object with spacetime indices by multiplication with tetrad components $Z^\mu{}_\nu = Z^a{}_b e^{b}{}_{\nu} e_{a}{}^{\mu}$. Spacetime indices are raised and lowered with the spacetime metric, tetrad indices are raised and lowered with the Minkowski metric.

\section{Coupling electromagnetic fields with torsion on Riemann-Cartan spacetime}\label{secelectro}
In this section we introduce Riemann-Cartan theories of gravity which include non-minimal couplings to the electromagnetic field strength tensor. 

The theory we are considering is built from the electromagnetic field strength tensors $F_{\mu\nu}$, the Riemann-Cartan curvature and the torsion tensor. As kinetic terms for the tetrad $e$ and the connection $\tilde \omega$ we introduce the Riemann-Cartan Ricci scalar and its square, while for the electromagnetic vector potential $A$ we introduce the usual Maxwell term, leading to:
\begin{align}
    \mathcal{L}_{\rm kinetic}[e,\tilde \omega, A]= -\tilde R - k_1 F_{\mu\nu}F^{\mu\nu} + k_2 \tilde{R}^2\,,\label{theoryA}
\end{align}
where the electromagnetic field strength is given in terms of the electromagnetic potential $1$-form $A$ as
\begin{equation}
    F_{\mu\nu}=2\partial_{[\mu}A_{\nu]}\,.
\end{equation}
\textbf{The term controlled by $k_2$ may effectively behave as a kinetic contribution for torsion 
(see~\eqref{eq:ECRicciS}) when suitable torsional invariants are incorporated.} 

The most general lowest order non-minimal coupling terms that are linear in the electromagnetic field strength $F$, linear in the curvature and at most quadratic in the torsion which can be constructed are
\begin{align}
    \mathcal{L}_{\rm int}[e,\tilde \omega, A] 
    &= k_3 F^{\mu  \nu  } \tilde{R}_{\mu  \nu  }+k_4 F^{\mu  \nu  } T_{\mu  }{}^{\alpha  \lambda  } T_{\alpha  \nu  \lambda  } + k_5 F^{\mu  \nu  } T_{\mu  \nu  }{}^{\alpha  } T_{\alpha  } + k_6 F^{\mu  \nu  } T^{\alpha  }{}_{\mu  \nu  } T_{\alpha  }\\
    &=:X_{\mu\nu}F^{\mu\nu}\,.\label{theoryB}
\end{align}
Applying integration by parts, this type of action can be rewritten as
\begin{align}
    \mathcal{L}_{\rm int}[e,\tilde \omega, A] =A_\mu j^\mu + 2\nabla^{[\mu}\Big(A^{\nu]}X_{\mu\nu}\Big)
\end{align}
with $j^\sigma  = g^{\sigma[\mu}\nabla^{\nu]}X_{\mu\nu}$. Since $X_{\mu\nu}$ is a pure geometric term, sourced by torsion, we can denote $j^\sigma$ as a source current for the electromagnetic field generated by torsion.

In the literature, also, additional quadratic terms in the torsion tensor (mass terms for the torsion), are often considered \cite{Bahamonde:2021gfp},
\begin{eqnarray}
    \mathcal{L}_{TT}[e,\tilde \omega]=k_7  T^{\mu\alpha  \rho  }T_{\mu  \alpha  \rho  }
    + k_8 T^{\mu \alpha  \rho  }T_{\alpha  \mu  \rho  }  
    +k_9 T_\alpha T^{\alpha  } \,.
\end{eqnarray}

The total Lagrangian, then takes the form
\begin{align}
    \mathcal{L}[e,\tilde \omega, A] 
    &= \mathcal{L}_{\rm kinetic} + \mathcal{L}_{\rm int}+\mathcal{L}_{TT}\\
    &= - R - k_1 F_{\mu\nu}F^{\mu\nu} + k_2 \tilde{R}^2  
  +k_3 F^{\mu  \nu  } \tilde{R}_{\mu  \nu  }  +\frac{1}{2}m_{T}^{2}T_{\mu}T^{\mu}+\frac{1}{2}m_{S}^{2}S_{\mu}S^{\mu}+\frac{1}{2}m_{t}^{2}t_{\lambda\mu\nu}t^{\lambda\mu\nu}\\
    &+ \frac{1}{2} k_4^{} \,\varepsilon_{\mu  \alpha  \lambda  \mu } F^{\nu  \alpha  } S^{\mu  } t^{\lambda  }{}_{\nu  }{}^{\mu } + (2 k_6^{}- k_5^{} ) F^{\nu  \alpha  } t_{\nu  \mu  \alpha  } T^{\mu  } + \frac{1}{6} (k_4^{} - k_5^{}  - k_6^{}) \,\varepsilon_{\mu  \nu  \alpha  \lambda  } F^{\alpha  \lambda  } S^{\mu  } T^{\nu  }\,,
\end{align}
where we introduced the irreducible pieces of the decomposition of torsion with respect to the 4-dimensional pseudo-orthogonal group as
\begin{align}
    T_{\mu}&=T^{\nu}\,_{\mu\nu}\,,\quad 
    S_{\mu}=\varepsilon_{\mu\lambda\rho\nu}T^{\lambda\rho\nu}\,,\quad 
    t_{\lambda\mu\nu}=T_{\lambda\mu\nu}-\frac{2}{3}g_{\lambda[\nu}T_{\mu]}-\frac{1}{6}\,\varepsilon_{\lambda\rho\mu\nu}S^{\rho}\,,\label{Tdec3}
\end{align}
and the mass parameters $m_t^2 =1 + 2 k_{7}^{} + k_{8}^{}, m_S^2 = \frac{1}{12} (1 - 4 k_{7}^{} + 4 k_{8}^{}) , m_T^2 = \frac{2}{3} (-2 + 2 k_{7}^{} + k_{8}^{} + 3 k_{9}^{})$ of the torsion fields, by using the expansion of $\tilde R$ from Eq.~\eqref{eq:ECRicciS}. We denote by $\varepsilon_{\mu \nu \alpha \lambda}$ the 4-dimensional Levi-Civita tensor. The torsion tensor in 4-dimensions can then be recovered from its irreducible pieces as
\begin{equation}\label{irreducibletorsion}
T^{\lambda}\,_{\mu \nu}=\frac{1}{3}\left(\delta^{\lambda}\,_{\nu}T_{\mu}-\delta^{\lambda}\,_{\mu}T_{\nu}\right)+\frac{1}{6}\,\varepsilon^{\lambda}\,_{\rho\mu\nu}S^{\rho}+t^{\lambda}\,_{\mu \nu}\,.
\end{equation}
While the direct torsion-field strength ($TTF$) couplings $k_4,k_5$ and $k_6$ have been introduced and studied for example in~\cite{Itin:2003hr}, a detailed study of the $\tilde R F$  field strength Riemann-Cartan curvature coupling is missing. We will provide this study in the following.

In order to isolate the effects that come from this type of non-minimal coupling, we reduce the theory to the Lagrangian
\begin{align} 
    \mathcal{L}[e,\tilde \omega, A]= -R- k_1F_{\mu\nu}F^{\mu\nu}+k_2 \tilde{R}^2+k_3 F^{\mu  \nu  } \tilde{R}_{\mu  \nu  }\,,\label{theory}
\end{align}
omitting the couplings $k_4,k_5,k_6$ and setting the masses of the torsion modes to zero. 

It is important to note that the coupling term $k_3 F_{\mu\nu} \tilde{R}^{\mu\nu}$ explicitly breaks the electromagnetic duality symmetry, since it is linear in the field strength $F$. In standard Maxwell theory in vacuum, the action is invariant under rotations of the electromagnetic field strength tensor of the form $F_{\mu\nu} \to \cos\theta \, F_{\mu\nu} + \sin\theta \, \ast F_{\mu\nu}$, where $\ast F_{\mu\nu} = \frac{1}{2} \varepsilon_{\mu\nu\alpha\beta} F^{\alpha\beta}$ is the dual field strength tensor. This symmetry reflects the equivalence of electric and magnetic sectors in the absence of charges and currents. However, the term $F_{\mu\nu} \tilde{R}^{\mu\nu}$ is linear in $F_{\mu\nu}$ it does not remain invariant under such rotations: under a duality transformation, it transforms into a combination involving $\epsilon_{\mu\nu\alpha\beta} F^{\alpha\beta} \tilde{R}^{\mu\nu}$, which is generically non-vanishing and distinct from the original term. This explicitly breaks the duality symmetry and leads to a theory where electric and magnetic sectors are no longer treated equivalently\footnote{This feature appears for any coupling for the electromagnetic field $A_\mu J^\mu$ to a current of the form $J^\mu = g^{\mu[\lambda}\nabla^{\sigma]}X_{\lambda\sigma}$, which is equivalent to a term $F_{\mu\nu}X^{\mu\nu}$.} in the action.

\textbf{It is worth emphasising that, in a strict effective-field-theory (EFT) sense, 
the torsion--electromagnetism coupling introduced in Eq.~\eqref{theory} is not mandated solely by diffeomorphism and $U(1)$ invariance. 
Our goal here, however, is not to construct a complete EFT expansion, but rather to explore a specific geometric realisation
of non-minimal couplings within the Riemann--Cartan framework. 
This point of view follows a long-standing tradition in torsion electrodynamics 
(see, e.g.,~\cite{Itin:2003hr,Hehl:1999bt,deAndrade:1997cj,Balakin:2007am,Solanki:2004az,Hammond:1991ty,Delhom:2020hkb,Dias2005,Hehl:2016wwp,shapiro2002,HehlObukhov2003}), 
where the emphasis is placed on covariant and geometrically motivated operators, rather than on EFT completeness. 
Accordingly, the operators in Eq.~\eqref{theory} should be regarded as a representative, geometrically natural subset of all possible non-minimal interactions, 
capturing the essential phenomenology of the torsionful electromagnetic sector.}

The field equations of the theory are obtained straightforwardly through variation of the action \eqref{theory} with respect to the fundamental variables. They can be expressed in terms of the tensorial quantities we introduced in the previous Sec.~\ref{sec:conventions}.

Variation with respect to the tetrads, and converting the indices to spacetime indices, yields
\begin{align}
   0 = E_{\mu\nu}=&\,G_{\mu\nu}+\frac{1}{2}k_1\Big(4F^\lambda{}_{\mu}F_{\lambda\nu}-F_{\sigma\lambda}F^{\sigma\lambda}g_{\mu\nu}\Big)+\frac{1}{2}k_2\tilde{R}\Big(\tilde{R}g_{\mu\nu}-4\tilde{R}_{\mu\nu}\Big)\nonumber\\
   &+\frac{1}{2}k_3\Big(F_{\lambda\rho}\tilde{R}^{\lambda\rho}g_{\mu\nu}-F^\lambda{}_{\mu}\tilde{R}_{\lambda\nu}-F^\lambda{}_\nu\tilde{R}_{\lambda\mu}+F^\lambda{}_\nu\tilde{R}_{\mu\lambda}-F^{\lambda\rho}\tilde{R}_{\mu\lambda\nu\rho}\Big)
\,.\label{FieldEq1}
\end{align}
To analyse these equations we split them conveniently into symmetric
\begin{align}
0=&\,G_{\mu\nu}+\frac{1}{2}k_1\Big(4F^\lambda{}_{\mu}F_{\lambda\nu}-F_{\sigma\lambda}F^{\sigma\lambda}g_{\mu\nu}\Big)+\frac{1}{2}k_2\tilde{R}\Big(\tilde{R}g_{\mu\nu}-4\tilde{R}_{(\mu\nu)}\Big)\nonumber\\
   &+\frac{1}{2}k_3\Big(F_{\lambda\rho}\tilde{R}^{\lambda\rho}g_{\mu\nu}-2\tilde{R}_{\lambda(\nu}F^\lambda{}_{\mu)}+F^\lambda{}_{(\nu}\tilde{R}_{\mu)\lambda}-F^{\lambda\rho}\tilde{R}_{(\mu|\lambda|\nu)\rho}\Big)
\end{align}
and antisymmetric parts
\begin{eqnarray}
   0= k_2 \tilde{R}_{[\mu  \nu ] } \tilde{R}+ \frac{1}{4} k_3{} \Big(F^{\lambda  \rho  } \tilde{R}_{[\mu | \lambda|  \nu]  \rho  } + \tilde{R}_{[\mu  }{}^{\lambda  }F_{\nu ] \lambda  } \Big)  \,.
\end{eqnarray}
Furthermore, by taking the trace of the tetrad equation~\eqref{FieldEq1} one finds that the theory in 4 dimensions always respect that the Levi-Civita Ricci scalar $R$ vanishes. Note that the symmetric part corresponds to the metric field equations.

Varying the action \eqref{theory} with respect to the spin connection, and converting all indices to spacetime indices, we arrive at the following connection field equations:
\begin{align}
  0= E^{\lambda \mu\nu} &=k_3\left(g^{\nu[\lambda}\nabla_{\rho}F^{\mu]\rho}-\nabla^{[\lambda}F^{\mu]}{}^{\nu}+g^{\nu [\lambda}K^{\mu]}{}_{\sigma\rho}F^{\sigma\rho}+K^{\rho[\lambda}{}_{\rho}F^{\mu]\nu}-K^{\nu[\lambda}{}_{\rho}F^{\mu]\rho}-K_{\beta\rho}{}^{[\lambda}g^{\mu]\beta}F^{\rho\nu}\right) \nonumber\\
   &+2k_2\Big(2\tilde{R}T^{\alpha[\lambda}{}_\alpha g^{\mu]\nu}-2\nabla^{[\lambda}\tilde{R}g^{\mu]\nu}-\tilde{R}T^{\nu\lambda\mu} \Big)\,.\label{FieldEq2}
\end{align}
Finally, variation with respect to the electromagnetic potential yields the generalized Maxwell equations
\begin{eqnarray}
   2k_1 \nabla_\mu F^{\mu\nu}&=&k_3\nabla_\mu\tilde{R}^{[\mu\nu]}\,.
   \label{FieldEq3}
\end{eqnarray}
As in standard Maxwell theory, it is $U(1)$ gauge invariant and we can fix the gauge to the Lorenz gauge $\nabla_\mu A^\mu=0$ to display this equation as
\begin{align}
    2k_1 (\Box A^\nu  + R^\mu{}_{\rho\mu\nu}A^\rho)&=k_3\nabla_\mu\tilde{R}^{[\mu\nu]} = j^\nu
\end{align}
with the torsion source being mediated by the coupling constant $k_3$.

To understand the impact of the $\tilde R F$ coupling we solve the field equations in spherical symmetry in 4-dimensions in the next section and in axial symmetry in 3-dimensions in Sec.\ref{sec:3D}.

\newpage
\section{Exact spherically symmetric black hole solutions in 4 dimensions}\label{sec:4D}
In this section, we will find 4-dimensional static and spherically symmetric solutions to the field equation \eqref{FieldEq1}, \eqref{FieldEq2} and \eqref{FieldEq3}, which we derived in the previous section, to understand the impact of the non-minimal coupling between the Riemann-Cartan Ricci tensor and the electromagnetic field strength. We distinguish two cases: $k_2=0$ in section \ref{ssec:k20} and $k_2\neq0$ in section \ref{ssec:k2}. In the next section \ref{sec:3D} we will continue this study in 3-dimensions.

We assume that the metric $g$, the electromagnetic potential $A$ and torsion tensor $T$ respect spherical symmetry and staticity, leading to
\begin{align}\label{sph_metric}
    g = ds^{2}&=\Psi_{1}(r)dt^{2} -\frac{dr^{2}}{\Psi_{2}(r)}-r^{2}d\vartheta^{2}-r^{2}\sin^{2}\vartheta \, d\varphi^{2}\,,\\
    A &= A_0(r) dt + A_r(r) dr\,,
\end{align}
and
\begin{align}\label{sph_torsion}
    T^t\,_{t r} &= t_{1}(r)\,, \quad T^r\,_{t r} = t_{2}(r)\,, \quad T^\vartheta\,_{t \vartheta} = T^\varphi\,_{t \varphi} = t_{3}(r)\,, \quad T^\vartheta\,_{r \vartheta} = T^\varphi\,_{r \varphi} =  t_{4}(r)\,, \\
    T^\vartheta\,_{t \varphi} &= T^\varphi\,_{\vartheta t} \sin^{2}\vartheta = t_{5}(r) \sin{\vartheta}\,, \quad T^\vartheta\,_{r \varphi} = T^\varphi\,_{\vartheta r} \sin^{2}\vartheta = t_{6}(r) \sin{\vartheta}\,, \\
    T^t\,_{\vartheta \varphi} &= t_{7}(r) \sin\vartheta\,, \quad T^r\,_{\vartheta \varphi} = t_{8}(r) \sin \vartheta\,.\label{sph_torsion2}
\end{align}
Due to the $U(1)$ gauge invariance of $F_{\mu\nu}$ we can choose $A_r=0$ without losing generality. Then, $A_0(r)$ is the only non-vanishing component of the electromagnetic field. In addition, there are two non-vanishing degrees of freedom coming from the metric ($\Psi_1(r),\Psi_2(r)$), and eight non-vanishing degrees of freedom from the torsion tensor $t_i(r)$.

\subsection{Solution without an additional kinetic term for torsion}\label{ssec:k20}
Assuming that $k_2=0$, meaning there is no additional kinetic term for the degrees of freedom sourcing torsion, and imposing spherical symmetry and staticity in the theory~\eqref{theory}, we find a unique solution of the connection \eqref{FieldEq2} and electromagnetic field \eqref{FieldEq3} equations. They are given by the following values of the components of torsion and the electromagnetic potential
\begin{align}
  t_1(r)&=-t_4(r)=\frac{1}{r}\,,\quad t_2(r)= t_3(r)= t_5(r)= t_6(r)= t_7(r)= t_8(r)=0\,,\quad A_0=\frac{q}{r}\,.
\end{align}
They have the property that $\tilde R^{[\mu\nu]} = 0$.

Solving in addition the tetrad field equations \eqref{FieldEq1} yields
\begin{eqnarray}
    \Psi_1(r)=\Psi_2(r)=1-\frac{2m}{r}+\frac{k_1  q^2}{r^2}\,,
\end{eqnarray}
defining the Reissner–Nordstr\"om metric from \eqref{sph_metric}.

In conclusion, this means that without the term $k_2 \tilde{R}^2$, the theory leads to the same spacetime metric as the standard Einstein-Maxwell theory with an additional non-propagating torsion tensor.

\subsection{Solution with an additional kinetic term for torsion}\label{ssec:k2}
To find more interesting solutions of the theory \eqref{theory}, we assume $k_2\neq0$, i.e. non-vanishing kinetic terms for the degrees of freedom in the torsion tensor. Even in the static and spherically symmetric case we are considering, solving the field equations becomes considerably more involved.

First of all, we implicitly solve the $E^{trr}$ connection component (see~\eqref{FieldEq2}) providing the following differential equation for $t_4$:
\begin{align}
  \Psi _1 \Psi _2 t_4'  =&\frac{k_3 \left(r t_4+1\right) \Psi _1 \Psi _2 A_0'}{16 k_2 r t_3}+\frac{\Psi _1 \Psi _2 \left(r^2 t_1'-r^2 t_4{}^2-4 r t_4+2 r t_1 \left(r t_4+1\right)-1\right)}{2 r^2}+\frac{1}{8} \Big[8 t_2 t_3+4 t_3{}^2-\frac{t_7{}^2 \Psi _1{}^2}{r^4}\nonumber\\
    &+\frac{t_8{}^2 \Psi _1}{r^4 \Psi _2}+\frac{4 t_5 t_7 \Psi _1}{r^2}+\frac{4 t_6 t_8 \Psi _1}{r^2}+\frac{4 \Psi _1}{r^2}+2 t_1 \Psi _2 \Psi _1'-4 t_4 \Psi _2 \Psi _1'+2 t_1 \Psi _1 \Psi _2'-4 t_4 \Psi _1 \Psi _2'+\frac{\Psi _2 \Psi _1'{}^2}{\Psi _1}\nonumber\\
    &-\Psi _1' \Psi _2'-\frac{4 \Psi _2 \Psi _1'}{r}-\frac{4 \Psi _1 \Psi _2'}{r}-2 \Psi _2 \Psi _1''\Big]\,,\label{difft4}
\end{align}
where primes denote derivatives with respect to $r$. By substituting this equation into the connection components $E^{t\vartheta\varphi}$, $E^{t\vartheta\vartheta}$, $E^{r\vartheta\varphi}$ and $E^{\vartheta\varphi t}$, one finds three branches of solutions for these equations: one with non-dynamical torsion (meaning $\tilde R = 0$) for which the metric is just the Reissner–Nordstr\"om metric; one that yields
\begin{align}
t_2(r)&=\pm\sqrt{\Psi _1(r)\Psi _2(r)}  \left(t_4(r)-t_1(r)-\frac{A_0''(r)}{A_0'(r)}\right)\pm\frac{\Psi _2(r) \Psi _1'(r)-\Psi _1(r) \Psi _2'(r)}{2 \sqrt{\Psi _1(r)\Psi _2(r)}}\,,\\
t_3(r)&=\mp \frac{\left(r\, t_4(r)+1\right) \sqrt{\Psi _1(r)\Psi _2(r)}}{r}\,,\\
t_5(r)&=\frac{t_7(r) \Psi _1(r)}{2 r^2}\pm  t_6(r) \sqrt{\Psi _1(r)\Psi _2(r)} \,,\\
t_8(r)&=\pm t_7(r) \sqrt{\Psi _1(r)} \sqrt{\Psi_2(r)}\,,
\end{align}
and the third one turns out to lead to a subcase of the above equations. The plus and minus signs correspond to two distinct branches of solutions arising from a quadratic equation. Next, by substituting the $E_{tt}$ tetrad component into $E_{rr}$ in~\eqref{FieldEq1}, we find
\begin{eqnarray}
    \Psi_1=\Psi_2=\Psi\,,
\end{eqnarray}
which automatically restricts the form of the metric to be of the Reissner–Nordström type, since the trace of the tetrad equation implies that the metric always satisfies $R = 0$. Then, by using again~\eqref{difft4}, we find from $E^{trt}$ that the electromagnetic potential has the standard Coulombian form:
\begin{eqnarray}
    A_0=\frac{q}{r}\,,
\end{eqnarray}
with $q$ being the electric charge. Since for the Coulom potential $\nabla_\mu F^{\mu\nu} = 0$, the modified Maxwell equation~\eqref{FieldEq3} implies that the torsion source must vanish: $\nabla_\mu \tilde{R}^{[\mu\nu]} = 0$. Furthermore, by using the differential equation~\eqref{difft4}, there are only two equations reamining in the system~\eqref{FieldEq1}--\eqref{FieldEq3}:
\begin{eqnarray}
  16 k_2 k_3 r^2 \Psi  A_0' t_1'  &=&-16 k_2 k_3 A_0' \left(r \left(r t_1+2\right) \Psi '+\Psi -1\right)\pm  \left(k_3^2-32 k_1 k_2\right) r^2 A_0'{}^2\mp 32   k_2 \left(r \Psi '+  \Psi-1\right) \,,\label{eq1solve}\\
\frac{1}{2}r^2 \Psi'' &=&1-2 r \Psi '-\Psi \,.\label{eq2solve}
\end{eqnarray}
By solving Eq.~\eqref{eq2solve}, the form of the metric function behaves as $\Psi(r) = 1 - \frac{2m}{r} + \frac{Q}{r^2}$ where $Q$ is an integration constant that acts as effective charge. Substituting this form of $\Psi(r)$ into Eq.~\eqref{eq1solve}, one can solve for the torsional function $t_1(r)$. The resulting expression is given by:
\begin{eqnarray}\label{eq:t1}
    t_1(r)
    &=&\frac{c_1}{\Psi (r)}\pm \frac{  \kappa _s}{r \Psi (r)}+\frac{\Psi '(r)}{2 \Psi (r)}\,
    =\frac{1}{\Psi (r)} \left(c_1\pm \frac{  \kappa _s}{r }+\frac{\Psi '(r)}{2 }\right)\,,
\end{eqnarray}
where $c_1$ and $\kappa_s$ are constants that absorb combinations of the parameters in the theory. In particular, the intrinsic spin charge $\kappa_s$ is identified through the term $1/(r \Psi(r))$, since the axial component of the torsion, $S_\mu$, behaves like a Coulomb potential that determines the torsion tensor (see also~\cite{Bahamonde:2020fnq}), and is given by, 
\begin{align}
    \kappa_s = \frac{2 Q}{k_3 q} + \frac{k_3 q}{16 k_2} - \frac{2 k_1 q}{k_3}\,.\label{kappas}
\end{align}
It is easy to see that if $k_3=0$ and $k_2\neq0$,  Eq.~\eqref{eq1solve} gives us $Q=k_1 q^2$ as in the standard Einstein-Maxwell case. Therefore, it is more natural to use~\eqref{kappas} to replace the general charge $Q$ in terms of the theory’s parameters, leading to the form for $\Psi(r)$ in terms of the theory parameters:
\begin{eqnarray}
    \Psi(r)
    &=1-\displaystyle\frac{2 m}{r}+\displaystyle\frac{1}{r^2}\left(k_1 q^2 - \frac{1}{2}k_3  \kappa_{\rm s} q - \frac{1}{32k_2}k_3 ^2 q^2\right)\label{eq:spinRN}\,.
\end{eqnarray}
As we assumed $k_2\neq0$ in our computations, the standard Einstein-Maxwell case would be recovered by imposing $k_3=0$.
This metric is of Reissner–Nordström form but exhibits a few important differences. The interaction between the spin charge $\kappa_{\rm s}$ and the electric charge $q$ can, depending on their magnitudes, change the sign of the term in brackets, which would completely alter the resulting phenomenology of the spacetime geometry. Further, the effective charge is not restricted to always be positive as in the standard Einstein-Maxwell case. Then, in principle, one can avoid the Cauchy horizon~\cite{Poisson:1989zz,Hollands:2019whz,Maeda:2005yd} with these corrections, and it has also been pointed out to occur in non-linear electrodynamics~\cite{Hale:2025ezt}.

Even though the Maxwell equations are unchanged in this solution, and thus the electromagnetic field does not directly feel the non-minimal coupling, it is affected indirectly through the spacetime metric, which clearly depends on the non-minimal coupling. In particular, because the coupling term $k_3 F_{\mu\nu} \tilde{R}^{\mu\nu}$ is linear in $F_{\mu\nu}$, it breaks the electromagnetic duality symmetry. This is already evident in our solution by the fact that the metric depends linearly on the electric charge $q$, distinguishing between positive and negative charges. While we have not constructed magnetic or dyonic configurations, it is expected that such solutions would not mirror the electric case, opening the possibility for qualitatively new phenomenology in the charged black hole sector. 

In addition to modifying the spacetime geometry, the intrinsic spin charge $\kappa_{\rm s}$ emerges as a new independent parameter of the black hole solution. In this sense, the solution provides a novel example of black hole ‘hair’ sourced by torsion, beyond the standard mass and electric charge parameters of Einstein--Maxwell theory. This raises interesting questions regarding the thermodynamics of the solution, and whether torsion contributes new conjugate potentials in the first law of black hole mechanics. Moreover, since torsion enters the gravitational sector indirectly but non-trivially, the presence of $\kappa_{\rm s}$ and the coupling $k_3$ are expected to affect the dynamics of perturbations. In particular, the spectrum of quasinormal modes, as well as the behavior of gravitational or electromagnetic perturbations around this solution, may differ from the standard Einstein--Maxwell case. A detailed perturbative analysis could thus reveal observable signatures of torsion, for instance through gravitational wave signals from black hole mergers or ringdown, providing new avenues to test these theories in astrophysical contexts. 

Notice that the remaining differential equation~\eqref{difft4} has two possible branches. The first occurs when $t_7(r) \neq 0$, in which case one can solve for $t_6(r)$ in algebraic form, leading to
\begin{eqnarray}
    t_6(r)&=&\frac{1}{t_7(r)}\Big[2 r t_4(r) \left(2-\frac{c_1 r}{\Psi (r)}\right)-\frac{4 c_1 r+1}{2 \Psi (r)}+r^2 t_4'(r)+r^2 t_4(r){}^2+2\Big]-\frac{t_7(r)}{4 r^2}\nonumber\\
    &&\mp \frac{1}{t_7(r) \Psi (r)}\Big[\frac{r^2 \left(r \Psi '(r)+\Psi (r)-1\right)}{k_3 q}+\frac{\left(k_3^2+32 k_1 k_2\right) q}{32 k_2 k_3}+2 r t_4(r) \kappa _s+2 \kappa _s\Big]\,.
\end{eqnarray}
Then, in this branch, $t_4(r)$ and $t_7(r) \neq 0$ remain arbitrary and thus parameterize the solution. On the other hand, the other branch, $t_7(r) = 0$, provides a differential equation for $t_4(r)$, explicitly given by
\begin{eqnarray}
   r^2 t_4'(r)&=& 2 r t_4(r) \left(\frac{c_1 r}{\Psi (r)}-2\right)+\frac{2 c_1 r+\frac{1}{2}}{\Psi (r)}-r^2 t_4(r){}^2+t_6(r)-2\nonumber\\
   &&\pm \frac{1}{\Psi(r)}\Big[\frac{r^2 \left(r \Psi '(r)+\Psi (r)-1\right)}{k_3 q}+\left(\frac{k_1}{k_3}+\frac{k_3}{32 k_2}\right) q+2 r t_4(r) \kappa _s+2 \kappa _s\Big]\,,
\end{eqnarray}
and the form of $t_6(r)$ remains arbitrary. Notice that all three irreducible parts of torsion described by~\eqref{Tdec3} are in general non-vanishing for our solution.

Before we turn to 3-dimensions we like to remark the following interesting observation on magnetic monopoles. Performing the analogous calculation as presented above for a purely magnetic potential $A_\mu = (0, 0, 0, -q_m \cos\theta)$, we can also solve the system, and the torsion and metric functions acquire imaginary contributions. Explicitly, the metric becomes $\Psi(r)=1-2m/r+(k_1 q_m^2-k_3^2 q_m^2-i k_3 q_m)/r^2$. This implies that the torsion field and the metric become complex-valued, leading to a complex spacetime geometry. Such behaviour is not expected in standard classical gravity, where both the metric and torsion are assumed to be real. This result may signal that the magnetic configuration is not compatible with real torsion in our model unless additional constraints are imposed.  One possibility is to require that the coupling $k_3$ vanishes in purely magnetic scenarios in order to ensure that all fields remain real. Alternatively, the theory might be interpreted as predicting the absence of magnetic monopoles in the gravitational sector. We leave a more detailed analysis of reality conditions and possible magnetic extensions for future work.

In the next section, we now turn to the three-dimensional case, where the non-minimal coupling between torsion and the electromagnetic field also leads to novel features, in particular in the context of deformations of the BTZ black hole.

\section{Exact black hole solutions in 3 dimensions}
\label{sec:3D}

Besides the interesting exact solution found in 4 dimensions, the non-minimal coupling between torsion and electromagnetism provides interesting features in three dimensions as deformation of the famous BTZ black hole~\cite{Banados:1992wn}. Three-dimensional black holes are important theoretical laboratories because gravity in three dimensions lacks local propagating degrees of freedom, allowing for a clearer analysis of the role played by global charges and topological effects. In particular, the BTZ black hole is a remarkable solution of (2+1)-dimensional gravity with a negative cosmological constant, featuring horizons, thermodynamic properties, and close connections to quantum gravity and holography, despite the absence of curvature singularities or gravitational radiation. Moreover, black hole solutions with torsion have been previously explored in various three-dimensional gravity models, such as Poincaré gauge theories and Chern--Simons-like formulations~\cite{Blagojevic:2013bu, Mielke:1991nn}. In these contexts, torsion can contribute non-trivially to conserved charges, modify the thermodynamics of the BTZ black hole, and influence the corresponding holographic dual field theories. In contrast to those models, where torsion typically plays the role of an effective cosmological constant, we will show below that in our model, torsion contributes instead as a charge-like parameter. Our model thus offers a new framework to further investigate the role of torsion in three-dimensional black hole physics. 

It turns out that a slowly rotating spin and electromagnetically charged BTZ black hole exhibits an interesting spin-angular-momentum coupling that is induced through by the electromagnetic charge and generated by the $k_3$ term in the action.

To perform the outline analysis in a systematic way, we add a negative cosmological term to the Lagrangian~\eqref{theory}, namely $L_{\Lambda}=-2\Lambda$ to include the standard BTZ solution~\cite{Banados:1992wn} as a limiting case, when the non-minimal couplings are set to zero. This extra term in the Lagrangian only introduces a term $-g_{\mu\nu}\Lambda$ in the tetrad equation~\eqref{FieldEq1}.

To find deformations of the BTZ black hole we consider a stationary and axially symmetric geometry, meaning using $\partial_t$ and $\partial_\phi$ as Killing vector fields: a 3-dimensional metric in coordinates $(t,r,\phi)$ of the form 
\begin{align}\label{sph_metric3D}
    ds^{2}&=(f(r)-r^2N_\phi(r)^2)dt^{2} -\frac{dr^{2}}{g(r)}-r^{2}d\phi^{2}-2r^2N_\phi(r)dt d\phi\,;
\end{align}
an electromagnetic potential of the form
\begin{align}
     A = A_t(r) dt + A_\phi(r) d\phi\,;
\end{align}
and a torsion tensor with all of its 9 independent components in 3 dimensions, parametrized as
\begin{align}
    T_{tt r} &= T_{1}(r)\,, \quad    T_{tt \phi} = T_{2}(r)\,, \quad  T_{tr\phi} = T_{3}(r)\,, \quad  T_{r t r} = T_{4}(r)\,, \quad   T_{r t \phi} = T_{5}(r)\,, \label{torsion3Da} \\
     T_{rr\phi} &= T_{6}(r)\,, \quad   T_{\phi t r} = T_{7}(r)\,,\quad   T_{\phi t \phi} = T_{8}(r) \,,\quad T_{\phi r \phi} = T_{9}(r)\,.\label{torsion3Dab}
\end{align}
Note that as occurs in the four-dimensional case, we can use the gauge freedom to eliminate any radial component of the potential, so we work in a gauge where $A_r = 0$. The remaining components $A_t(r)$ and $A_\varphi(r)$ capture the physical degrees of freedom of the electromagnetic field in this background.

We now first consider the exact non-rotating case, for which we find a charged BTZ-like solution, before we extend this solution to a charged slowly rotating BTZ like black hole.

\subsection{Spherically symmetric exact charged BTZ-like solution}\label{ssec:nonrtoBTZ}
To find a non-rotating charged BTZ-like black hole, that can serve as a background solution for the slowly rotating case, we employ the metric Ansatz \eqref{sph_metric3D} with $N_\varphi = 0$ and the electromagnetic component $A_\phi=0$.

The field equations \eqref{FieldEq1}-\eqref{FieldEq3} can be easily solved giving us $g(r)=f(r)$ and the torsion components become
\begin{eqnarray}
  T_{1}&=&f(r) \left(\pm T_{4}(r)+\frac{1}{r}\right)\,,\quad T_{2}=\pm \frac{1}{2} K_1 f(r)\,,\quad T_{3}= T_{5}=0\,,\quad T_{7}=K_1\,,\quad \label{eq:BTZTor1} \\
T_{4}&=&\pm \frac{f'(r)}{2 f(r)}\mp \frac{1}{r}+\frac{K_2}{f(r)}+\frac{\kappa _s\log r }{f(r)}\,,\quad T_{6}=\mp\frac{K_1}{2 f(r)}\,,\quad T_{8}=\pm f(r) \left(r-T_{9}(r)\right) \label{eq:BTZTor2}\,,
\end{eqnarray}
where primes denote derivatives with respect to $r$, $\pm$ denotes the two possible solutions and $K_1,K_2$ and $\kappa_{\rm s}$ are integration constants. The last torsional degree of freedom $T_{9}$ satisfies the following differential equation
\begin{eqnarray}
  T_{9}'&=&\frac{K_1^2}{2 f}-1+\frac{f' \left(-k_3 q_e T_{9}+k_3 r q_e\pm r^2\right)}{k_3 q_e f}\mp 2 T_{4} \left(r-T_{9}\right)+\frac{2 T_{9}}{r}\pm\frac{2 \Lambda  r^3}{k_3 q_e f}\nonumber\\
  &&\pm\frac{k_3 r q_e}{16 k_2 f}\pm \frac{2 k_1 r q_e}{k_3 f}\,. \label{eq:BTZTor3}
\end{eqnarray}
Similarly, as we noted in the previous Sec.~\ref{sec:4D} in 4-dimensions, the constant $\kappa_{\rm s}$ has the interpretation of an intrinsic spin charge. 

For the electromagnetic potential, we find the logarithmic behaviour
\begin{eqnarray}\label{eq:BTZA}
    A_{t}=q_e \log r\,,
\end{eqnarray}
where $q_e$ is the electric charge. Finally, the metric solution is an extension of the non-rotating charged BTZ black hole in 3-dimensions \cite{Martinez:1999qi}, with a new coupling related to the intrinsic spin mediated by the new interaction term $\kappa_{\rm s} k_3 q_e$:
\begin{eqnarray}\label{eq:BTZmet}
    g(r)=f(r)=-M-\Lambda  r^2- \Big(2 k_1 q_e^2-\frac{k_3^2 q_e^2}{16 k_2}+\kappa_{\rm s} k_3 q_e\Big)\log r\,,\quad N_\phi = 0\,.
\end{eqnarray}
Indeed, this solution shows a similar behavior to the 4-dimensional case in that both geometries acquire modifications to the black hole metric sourced by the intrinsic spin charge $\kappa_s$ through the non-minimal coupling $k_3 F^{\mu\nu} \tilde{R}_{\mu\nu}$. In both cases, the electric charge couples linearly to torsion, resulting in an effective spin-charge interaction that alters the black hole structure. While the specific radial dependence of the corrections differs, with $1/r^2$ corrections in four dimensions and $\log r$ terms in three, the underlying mechanism of torsion-induced spin-charge effects is common to both, marking a consistent feature of the theory across dimensions.

Compared to previously known three-dimensional solutions with torsion, where torsion typically appears as an effective constant or as a modification of the cosmological term~\cite{Mielke:2003xx,Garcia:2003nm,Blagojevic:2003wn}, the new feature our solution exhibits is dynamically coupled torsion components that evolve with the radial coordinate. Moreover, it gives rise to an intrinsic spin-charge hair in three dimensions and then, the thermodynamic behaviour of the black hole may be modified with respect to the non-rotating charged BTZ solution.

Having found a non-rotating spin-charged extension of the BTZ black hole, we can consider slowly rotating ones, to find new couplings between elementary and angular-momentum spin.

\subsection{Slowly rotating charged BTZ-like solution}
To find slowly rotating solutions, meaning small angular momentum parameter $J$, we assume that all fields $X$ can be expanded in a background solutions $X{}_A$, given by the solution found in Section \ref{ssec:nonrtoBTZ}, displayed in equations \eqref{eq:BTZTor1}-\eqref{eq:BTZTor3}, \eqref{eq:BTZA} and \eqref{eq:BTZmet}, and a correction $X_B$, that is multiplied linearly with $J$, namely,
\begin{eqnarray}
    T_i(r)&=&T_{i}(r)+J\, T_{i,B}(r) \,,\quad A_t(r)=A_{t}(r)+J\, A_{t,B}(r)\,,\quad  A_\phi(r)=J\, A_{\phi,B}(r)\,,\\
   f(r)&=&f_{A}(r)+J f_{B}(r) \,,\quad  g(r)=f_{A}(r)+J g_{B}(r) \,,\quad N_\phi=J\, N_{\phi,B}\quad |J| \ll 1\,.
\end{eqnarray}
Notice that in the previous section we found that $g(r)=f(r)$, and then, for simplicity, we will assume that this condition holds for the slowly rotating case, i.e., $g_{B}(r)=f_{B}(r)$. Expanding our field equations~\eqref{FieldEq1}, \eqref{FieldEq2} and~\eqref{FieldEq3} up to first order in the angular momentum $J$ allows us to solve for the fields $X_B$. 

The computation is quite involved, but we will provide the main steps here. First, we solve the connection field component $E^{rt\varphi}$ (see~\eqref{FieldEq3}) for $T'_{9,B}$ and replace that expression into the other field components. Then, one can algebraically solve $E^{\phi r t},E^{t \phi r},E^{\phi rr},E^{\phi tt },E^{\phi t \phi}$ yielding
\begin{eqnarray}
    T_{1,B}&=&f_B(r) \left(\frac{1}{r}\pm  T_{4,A}(r)\right)+\pm  f_A(r) T_{4,B}(r)+\pm  H_1(r) f_A(r)+\frac{1}{2} K_1 N_{\phi ,B}\,,\label{T1B}\\
      T_{2,B}&=&-\frac{f_A(r){}^2 A_{\phi ,B}'(r)}{2 r q_e}\Big(r^2 T_{4,A}(r)\pm  \left(T_{9,A}(r)+r\right)\Big) \pm f_A(r) \left(  H_2(r)-\frac{  K_1 H_1(r)}{2 T_{4,A}(r)}\right) \pm\frac{1}{2}  K_1 f_B(r)\,,\\
      T_{3,B}&=&T_{9,A}(r) \left(\frac{f_A(r) A_{\phi ,B}'(r)}{r q_e}+N_{\phi ,B}\right)-\frac{  f_A(r) A_{\phi ,B}'(r) }{q_e}\left( 1\pm r T_{4,A}(r) \right)\,,\\
          T_{5,B}&=&0\,,\\
            T_{6,B}&=&\mp\frac{ \left(T_{9,A}(r)+r\right) A_{\phi ,B}'(r)}{2 r q_e}-\frac{r T_{4,A}(r) A_{\phi ,B}'(r)}{2 q_e}\pm \frac{  K_1 f_B(r)}{2 f_A(r){}^2}\mp\frac{3   K_1 H_1(r)}{2 f_A(r)T_{4,A}(r)}\pm \frac{3   H_2(r)}{f_A(r)}\nonumber\\
            &&\pm\frac{1}{f_A(r)}\left(\left(T_{9,A}(r)-3 r\right) N_{\phi ,B}(r)-\frac{r f_A'(r) A_{\phi ,B}'(r)+r^2 q_e N_{\phi ,B}'(r)+2 q_e T_{7,B}(r)}{q_e}\right)\,,\\
              T_{8,B}&=&\mp\frac{ f_A(r){}^2 \left(r-T_{9,A}(r)\right) }{2 K_1 q_e}\left(r A_{\phi ,B}''(r)-A_{\phi ,B}'(r)\right)\pm  f_B(r) \left(r-T_{9,A}(r)\right)\nonumber\\
              &&\mp  f_A(r) \left(\frac{r \left(r q_e \left(r-T_{9,A}(r)\right) N_{\phi ,B}'(r)+A_{\phi ,B}'(r) \left(\left(r-T_{9,A}(r)\right) f_A'(r)+K_1^2\right)\right)}{2 K_1 q_e}+T_{9,B}(r)\right)\,,\label{T8B}
\end{eqnarray}
where we have introduced
\begin{eqnarray}
    H_1(r)&=&\pm \frac{  K_1 A_{\phi ,B}'(r)}{2 r q_e}+\frac{T_{4,A}(r)}{2 K_1 q_e} \Big[r \left(f_A'(r) A_{\phi ,B}'(r)+r q_e N_{\phi ,B}'(r)\right)+f_A(r) \left(r A_{\phi ,B}''(r)-A_{\phi ,B}'(r)\right)\Big]\nonumber\\
    &&\mp\frac{  \left(A_{t,B}'(r)+r A_{t,B}''(r)\right)}{q_e}\,,\label{H1}\\
     H_2(r)&=&\frac{1}{2} \left(r \left(\frac{f_A'(r) A_{\phi ,B}'(r)}{q_e}+r N_{\phi ,B}'(r)\right)+\left(2 r-T_{9,A}(r)\right) N_{\phi ,B}(r)+T_{7,B}(r)\right)\nonumber\\
     &&\pm \frac{K_1}{2q_eT_{4,A}(r)}\Big[\frac{  K_1 A_{\phi ,B}'(r)}{2 r }-   \left(A_{t,B}'(r)+r A_{t,B}''(r)\right)\Big]\,.\label{H2}
\end{eqnarray}
Next, we use $E^{rt\varphi}, E^{\phi r \phi}$ and~$E^{r tt}$ to find $T'_{9,B},N_{\phi,B}''$ and $A_{t,B}''$ and replace them (and their derivatives) into the tetrad field equation~\eqref{FieldEq1} components $E_{\phi t},E_{\phi \phi}, E_{tt}$ and $E_{rr}$. Then, by combining those equations in a nontrivial way, we find the following simple expressions:
\begin{eqnarray}
    f''_B(r)&=&-\frac{1}{r}f_B'(r)\,,\quad  A_{t,B}''(r)= -\frac{A_{t,B}'(r)}{r}\,,\label{eqimportant1}\\
    N'''_{\phi ,B}(r)&=&-\frac{\left(r \left(5 f_A'(r)+4 \Lambda  r\right)+3 f_A(r)\right) N_{\phi ,B}'(r)}{r^2 f_A(r)}-\frac{\left(r f_A'(r)+5 f_A(r)\right) N_{\phi ,B}''(r)}{r f_A(r)}\,,\\
    A_{\phi ,B}''(r)&=& \frac{A_{\phi ,B}'(r)}{r}-\frac{f_A'(r) A_{\phi ,B}'(r)+r q_e N_{\phi ,B}'(r)}{f_A(r)}\,,
\end{eqnarray}
that from~\eqref{H1} implies $H_1(t)=\pm \tfrac{  K_1 A_{\phi ,B}'(r)}{2 r q_e}$. The first two equations provide a logarithmic form for $f_B$ and $A_{t,B}$ and the last two equation can be solved by setting
\begin{eqnarray}
    N_{\phi,B}(r)=-K_3-\frac{q_m}{q_e r^2}f_A(r)\,,\quad A_{\phi ,B}(r)= q_m \log r+K_4\int \frac{r dr}{f_A(r)}  \,,\label{eqimportant}
\end{eqnarray}
where $K_3,K_4$ and $q_m$ are integration constants with the latter being a magnetic charge.

By replacing these solutions into the torsional functions~\eqref{T1B}-\eqref{T8B} we find
\begin{eqnarray}
    T_{1,B}&=&f_B(r) \left(\frac{1}{r}\pm  T_{4,A}(r)\right)\pm  f_A(r) T_{4,B}(r)+\frac{K_1 K_4}{2 q_e}-\frac{1}{2} K_1 K_3\,,\\
    T_{2,B}&=& \pm\frac{1}{2}  f_A(r) \left(K_3 T_{9,A}(r)+T_{7,B}(r)-2 K_3 r\right)-\frac{q_m f_A(r){}^2 }{2 q_e}\left(T_{4,A}(r)\pm \frac{1 }{r}\right)\pm \frac{1}{2}   K_1 f_B(r)\nonumber\\
    &&-\frac{K_4}{2q_e}\Big[r^2 f_A(r) T_{4,A}(r)\pm  \left(f_A(r) \left(T_{9,A}(r)+r\right)-r^2 f_A'(r)\right)\Big]\,,\\
    T_{3,B}&=&-\frac{q_m f_A(r) }{q_e}\left(\frac{1}{r}\pm  T_{4,A}(r)\right)-\frac{K_4 }{q_e}\left(r-T_{9,A}(r)\pm r^2 T_{4,A}(r)\right)-K_3 T_{9,A}(r) \,,\quad T_{5,B}=0\,, \\
     T_{6,B}&=&\pm\frac{  \left(K_3 T_{9,A}(r)-T_{7,B}(r)\right)}{2 f_A(r)}-\frac{K_4 }{2 q_e f_A(r)}\left(r^2 T_{4,A}(r)\pm  \left(T_{9,A}(r)-\frac{r^2 f_A'(r)}{f_A(r)}+r\right)\right)\nonumber\\
     &&+\frac{q_m \left(-r T_{4,A}(r)\pm1 \right)}{2 r q_e}\pm\frac{  K_1 f_B(r)}{2 f_A(r){}^2}\,,\\
     T_{8,B}&=&\pm  f_A(r) \left(-T_{9,B}(r)-\frac{K_1 q_m}{2 q_e}\right)\pm  f_B(r) \left(r-T_{9,A}(r)\right)\mp \frac{  K_1 K_4 r^2}{2 q_e}\,,
\end{eqnarray}
with the remaining torsional functions satisfying the following differential equations:
\begin{eqnarray}
 T_{9,B}'&=&f_A' \left(\frac{q_e f_B \left(T_{9,A}-r\right)-f_A \left(q_e T_{9,B}+K_1 q_m\right)}{q_e f_A{}^2}\mp\frac{  r^3 A_{t,B}'}{k_3 q_e^2 f_A}\mp \frac{  r^2 f_B}{k_3 q_e f_A{}^2}\right)+\frac{K_1 \left(T_{7,B}-K_3 r\right)}{f_A}\nonumber\\
 &&+\frac{2 q_e T_{9,B}+K_1 q_m}{r q_e}\pm \frac{  r^2 \left(k_3^2 q_e^2+32 k_1 k_2 q_e^2-32 k_2 \Lambda  r^2\right) A_{t,B}'}{16 k_2 k_3 q_e^2 f_A}+f_B' \left(\frac{r-T_{9,A}}{f_A}\pm \frac{  r^2}{k_3 q_e f_A}\right)\nonumber\\
 &&\pm   \left(2 \left(T_{9,A}-r\right) T_{4,B}-\frac{r f_B \left(k_3^2 q_e^2+32 k_1 k_2 q_e^2+32 k_2 \Lambda  r^2\right)}{16 k_2 k_3 q_e f_A{}^2}\right)-\frac{K_1^2 f_B}{2 f_A{}^2}\nonumber\\
 &&\pm   T_{4,A} \left(2 T_{9,B}+\frac{K_1 K_4 r^2}{q_e f_A}+\frac{K_1 q_m}{q_e}\right)\,,\\ 
16 k_2 k_3 r^2 q_e^2 f_A{}^2 T_{4,B}' &=&f_A \Big[-16 k_2 k_3 r^2 q_e^2 T_{4,A} f_B'+\left(k_3^2-32 k_1 k_2\right) r^2 q_e^2 A_{t,B}'+32 k_2 \Lambda  r^4 A_{t,B}'\nonumber\\
&&-16 k_2 r^2 q_e f_B'\mp 24   k_2 k_3 r q_e^2 f_B'\Big]+32 k_2 \Lambda  r^3 q_e f_B\pm 32   k_2 k_3 \Lambda  r^2 q_e^2 f_B+\left(32 k_1 k_2-k_3^2\right) r q_e^3 f_B\nonumber\\
&&+f_A' \Big[16 k_2 k_3 r^2 q_e^2 f_B T_{4,A}+f_A \left(16 k_2 r^3 A_{t,B}'-16 k_2 k_3 r^2 q_e^2 T_{4,B}\right)\nonumber\\
&&+16 k_2 r^2 q_e f_B\pm 24  k_2 k_3 r q_e^2 f_B\Big]\,,\\
 16 k_2 k_3 r q_e^2 f_A T_{7,B}'&=&f_A' \Big[16 k_2 k_3 K_3 r q_e^2 \left(r-T_{9,A}\right)+k_2 q_e \left(\pm 16 K_3 r^3-8 k_3 q_m f_A\right)\mp 16 k_2 r q_m f_A\Big]\nonumber\\
 &&+q_e^2 \Big[k_2 \Big\{k_3 \left(32\pm K_3 r f_A T_{4,A} \left(T_{9,A}-r\right)-16 K_3 f_A \left(r-2 T_{9,A}\right)+8 K_3 K_1^2 r\right)\mp 32 k_1 q_m f_A\Big\}\nonumber\\
 &&\pm k_3^2 q_m f_A\Big]+f_A' \Big[\pm 16   k_2 K_4 r^3 \left(k_3 q_e T_{4,A}+2\right)+8 k_2 k_3 K_4 r q_e \left(2 T_{9,A}+r\right)\Big]\nonumber\\
 &&+k_2 q_e \left(\pm 32 \Lambda  K_3 r^4-32 k_3 \Lambda  r q_m f_A\right)\mp 32 k_2 \Lambda  r^2 q_m f_A\pm q_e^3 \left(k_3^2 K_3 r^2+32 k_1 k_2 K_3 r^2\right)\nonumber\\
 &&\mp 32  k_2 K_4 r \left(k_3 q_e f_A T_{4,A} T_{9,A}+2 k_1 r q_e^2-2 \Lambda  r^3\right)\nonumber\\
 &&-8 k_2 k_3 K_4 q_e \left(4 f_A T_{9,A}-2 r f_A+K_1^2 r-4 \Lambda  r^3\right)\,.
\end{eqnarray}
Using~\eqref{eqimportant1} and \eqref{eqimportant}, we find that the final form of the vector field up to first order in $J$ is
\begin{eqnarray}
    A_\mu=\Big(q_e\log r+J K_5\log r\,,0\,,J q_m \log r+J K_4\int \frac{r dr}{f_A(r)}\Big)\,,
\end{eqnarray}
where $K_5$ is another integration constant. The final form of the metric is:
\begin{align}\label{sph_metric3DFinal}
    ds^{2}&=\Big(f_{A}(r)+J f_{B}(r)\Big)dt^{2} -\Big(\frac{1}{f_{A}(r)}+J\,\frac{f_B(r)}{f_{A}^2}\Big)dr^{2}-r^{2}d\phi^{2}+2\,\Big(K_3\,J\, r^2+\,J\,\frac{q_mf_{A}(r)}{q_e}\Big)dt d\phi\,,
    \end{align}
    where
\begin{eqnarray}
    f_{A}(r)&=&-M-\Lambda  r^2- \Big(2 k_1 q_e^2-\frac{k_3^2 q_e^2}{16 k_2}+\kappa_{\rm s} k_3 q_e\Big)\log r\,,\\
    f_{B}(r)&=&-K_6-K_7 \log r\,,
\end{eqnarray}
where we have solved~\eqref{eqimportant1} for $f_B$, and $K_6$ and $K_7$ are integration constants.
We notice a new coupling between the intrinsic spin $\kappa_{\rm s}$, the angular momentum of the rotation $J$ and the magnetic charge $q_m$, i.e. $\propto J q_m \kappa_{\rm s} $. This new interaction, is mediated by the new term in the Lagrangian that is related to $k_3$. Interestingly, the solution also displays an asymmetric dependence on the electric and magnetic charges due to the non-minimal coupling term, which leads to modifications in the near-horizon behaviour. A feature that could potentially influence the black hole entropy and the first law of black hole thermodynamics, when angular momentum is present, as the usual identification between surface gravity, angular velocity, and temperature may no longer hold without torsion corrections. 

In the context of the AdS/CFT correspondence, BTZ black holes play a central role as gravity duals of thermal states in two-dimensional conformal field theories (CFTs)~\cite{Brown:1986nw}. The presence of torsion, particularly when it is dynamically sourced and couples non-minimally to the electromagnetic field, suggests that the dual theory is no longer a standard CFT but rather a deformation of it, possibly one with spin current sources or broken parity.

In our model, torsion components are activated by both charge and rotation, and they backreact on the metric. This is expected to modify the boundary stress tensor and the holographic Ward identities, as it happens in~\cite{Klemm:2007yu,Blagojevic:2013bu}. In particular, the asymmetry between electric and magnetic sectors due to the coupling mediated by $k_3$ implies that the dual theory could distinguish between left- and right-movers, suggesting a chirally asymmetric dual CFT~\cite{Castro:2014tta,Detournay:2012pc}. Moreover, the spin-charge hair sourced by torsion is expected to affect boundary transport phenomena associated with geometry, such as Hall viscosity or thermal Hall response. In particular, torsion has been shown to couple to spin currents and induce dissipationless transport coefficients in topological phases of matter~\cite{Hughes:2011hv,Hughes:2012vg,Gromov:2014vla}. These effects suggest that the holographic dual theory may exhibit torsional anomalies or contain new geometric response terms beyond the usual energy-momentum and electromagnetic current structure.

Finally, since torsion modifies the thermodynamics and potentially the central charges (via boundary charges in Chern–Simons-like formulations), the Cardy formula may receive corrections as in~\cite{Park:2006gt,Blagojevic:2006jk,Blagojevic:2006hh}. This could affect the microscopic counting of black hole entropy in the dual theory, an aspect worth exploring in future work.

\section{Conclusions}\label{sec:conc}
Due to the two-form nature (the anti-symmetric index pair) of the torsion and the electromagnetic field strength tensor, it is natural to study multiple non-trivial, but natural, ways of how the torsion tensor can couple to the electromagnetic field strength tensor in Riemann-Cartan geometry. Among the many terms there are, we studied and focused on new interactions of the form $k_3F^{\mu\nu}\tilde{R}_{\mu\nu}$, which have not been studied in detail previously.

We found that, just adding this term to the usual Einstein-Hilbert action does not lead to any interesting Riemann-Cartan geometry solutions beyond GR. In order to find exact non-trivial black hole solutions in four and three dimensions beyond GR, we added a kinetic term for the torsion tensor, namely $k_2\tilde{R}^2$. We derived the general field equation for this class of theories and found in particular, that the Maxwell equations are sourced by the covariant divergence of the anti-symmetric part of the Riemann-Cartan Ricci tensor.

In order to find black hole solutions in 4-dimensions we assumed spherical symmetry for all involved fields. We found that in this case, the electromagnetic potential is the Coulomb potential, torsion is non-trivial, and the spacetime metric becomes a spin-charged generalisation of the Reissner–Nordström solution, which we presented in Eq.~\eqref{eq:spinRN}. The radial dependence of the metric in the new solution has the same form as that of the Reissner–Nordström solution, but with an effective electric charge. Novel features emerge from the total charge. In our new solution, this charge is not only given by the electric charge $q$, but by a combination of electric charge, the spin charge $\kappa_s$ and the coupling constants $k_2$ and $k_3$, which induce dynamical torsion and the non-minimal coupling between torsion and electromagnetism. The most intriguing feature of the solution is that, depending on the strength of the couplings and the size of the spin-charge, the total charge need not to be positive, which changes the phenomenology compared to the usual Reissner–Nordström solution. A future interesting study will be to investigate the phenomenology of these spin-charged black holes, for example the behaviour of their accretion discs, as extension of existing studies on spin-charged accretion discs around general relativistic black holes \cite{Gimeno-Soler:2025jnl,Lahiri:2023mwj}.

In 3-dimensions, we added a negative cosmological constant to the theory in order to find new torsion-based generalization of the BTZ black hole~\cite{Banados:1992wn}. As in 4-dimension, it turns out that the source term of the Maxwell equation, the covariant divergence of the anti-symmetric part of the Riemann-Cartan Ricci tensor vanishes and thus the Maxwell equations are unchanged. For the metric, we found an electromagnetically and spin charged non-rotating BTZ black hole solution \eqref{eq:BTZmet}, which is the 3-dimensional analogue to the previously discussed 4-dimensional solution. It serves as a background solution for our extension to a slowly rotating solution, \eqref{sph_metric3DFinal}. It exhibits a remarkable feature absent in the original BTZ solution: a spin-charge–angular momentum coupling induced by the non-minimal coupling constant $k_3$. This result highlights the possibility of richer torsional contributions in rotating geometries, motivating further exploration in various dimensions and potentially revealing new holographic properties.

In the cases we considered in detail in this article, we found that the Riemann-Cartan source term of the Maxwell equations $\nabla_\mu\tilde{R}^{[\mu\nu]}$ vanishes. For future investigations it would be interesting to find solutions where the Maxwell equations posses this torsion-based source term. One way to study this situation is to consider flat, torsion free Minkowski spacetime with vanishing electromagnetic potential as background spacetime and perturb this setup linearly by general torsion (24 dof), electromagnetic field (4 dof), and metric (10 dof). Keeping the perturbations general, not subject to any symmetry assumptions. This study might reveal features and metrics that differ from the form of the formal Reissner–Nordström solution.

When a non-trivial background for the metric is considered, such as Reissner-Nordström or Kerr-Newmann spacetime geometry without torsion and electromagnetic field, perturbations can be studied too determine torsional quasinormal that may provide new signatures of torsion, potentially detectable through observations of gravitational waves emitted by binary black hole mergers. Further, superradiance generated by our solution might trigger new effects related to torsion~\cite{Brito:2015oca}.

Further future directions include relaxing the constraint $k_4 = k_5 = k_6 = 0$ to uncover new black hole geometries with enricher torsional structures. On the cosmological side, it would be compelling to study solutions that incorporate torsion-induced modifications to the metric and electromagnetic sources, particularly in scenarios involving magnetic fields. It is well known that magnetic fields play an important role in the early universe~\cite{Grasso:2000wj}, and since our theory naturally extends Einstein–Maxwell with torsion, it would be interesting to analyse the evolution of primordial magnetic fields within this framework, with the aim of identifying new effects of torsion in the early universe. Another potentially significant direction is to investigate whether our theory can induce spontaneous parity violation at the level of linear perturbations, as commonly occurs in theories with torsion or nonmetricity~\cite{Hohmann:2022wrk,Aoki:2023sum}. This mechanism could offer a natural explanation for the cosmic birefringence reported in~\cite{Minami:2020odp}. Furthermore, extending the analysis of rotation to four-dimensional solutions could clarify whether the coupling between intrinsic spin charge and angular momentum—observed in the three-dimensional case—also arises in higher-dimensional rotating black holes. Such studies could uncover new thermodynamic and geometric features induced by torsion. We leave all of these studies for the future.

\textbf{Finally, it would be interesting to analyse if our theory (mainly the coupling controlled by $k_3$) can arise from a higher-dimensional Einstein-Cartan (or higher-derivative curvature) gravity including the topological invariants (Euler, Pontryagin/Chern, and Nieh--Yan) and then perform a Kaluza--Klein dimensional reduction. This kind of reduction is known to generate in four dimensions non-minimal electrodynamics with curvature--electromagnetism cross terms (e.g.\ $R\,F^2$, $R_{\mu\nu}F^{\mu\alpha}F^{\nu}{}_{\alpha}$, etc.), see for instance~\cite{MuellerHoissen:1988PLB,Salam:1981xd,DereliUcoluk:1990CQG,Cremmer:1978km}. Exploring whether an Einstein--Cartan parent theory with such topological densities can induce couplings of the schematic form $\tilde{R}^{\mu\nu}F_{\mu\nu}$ is a natural continuation of our work and will be addressed elsewhere.}

\section*{Acknowledgements}
S.B. is supported by “Agencia Nacional de Investigación y Desarrollo” (ANID), Grant “Becas Chile postdoctorado al extranjero” No. 74220006 and also by Institute for Basic Science (IBS-R018-D3). 
J.M. is supported by  ANID through a doctoral scholarship and internship funding (No. 21212072), as well as institutional support from UTFSM and PUCV. He also thanks the Kavli IPMU for hosting him as a visiting researcher.
C.P. acknowledges support by the excellence cluster QuantumFrontiers of the German Research Foundation (Deutsche Forschungsgemeinschaft, DFG) under Germany's Excellence Strategy -- EXC-2123 QuantumFrontiers -- 390837967 and was funded by the Deutsche Forschungsgemeinschaft (DFG, German Research Foundation) - Project Number 420243324.
The authors also like to thank Jorge Gigante Valcarcel and Theodoros Nakas for useful discussions.
\bibliographystyle{utphys}
\bibliography{references}

 
\end{document}